\documentclass[a4paper,fleqn,usenatbib]{mnras}

\usepackage{newtxtext,newtxmath}

\usepackage[T1]{fontenc}
\usepackage{ae,aecompl}

\usepackage{graphicx}	
\usepackage{amsmath}	
\usepackage{amssymb}	

\usepackage{multirow}
\usepackage{bm}

\newcommand{\dd}{\mathrm{d}}

\newcommand*{\be}{\begin{equation}}
\newcommand*{\ee}{\end{equation}}
\newcommand{\bea}{\begin{eqnarray}}
\newcommand{\eea}{\end{eqnarray}}

\title[Constraining primordial non-Gaussianity]{Constraining primordial non-Gaussianity 
using two galaxy surveys and CMB lensing}

\author[M.~Ballardini, W.~L.~Matthewson, R.~Maartens]{Mario Ballardini$^{1,2}$\thanks{Contact e-mail: \href{mailto:mario.ballardini@gmail.com}{mario.ballardini@gmail.com}}, 
William L. Matthewson$^{1}$\thanks{Contact e-mail: \href{mailto:willmatt4th@gmail.com}{willmatt4th@gmail.com}}, Roy Maartens$^{1,3}$\thanks{Contact e-mail: \href{mailto:roy.maartens@gmail.com}{roy.maartens@gmail.com}} 
\\
$^{1}$Department of Physics \& Astronomy, University of the Western Cape, Cape Town 7535, South Africa \\
$^{2}$INAF/OAS Bologna, via Gobetti 101, I-40129 Bologna, Italy \\
$^{3}$Institute of Cosmology \& Gravitation, University of Portsmouth, Portsmouth PO1 3FX, UK}

\pubyear{2019}

\begin{document}
\maketitle

\begin{abstract}
Next-generation galaxy surveys will be able to measure perturbations on scales beyond the equality scale. 
On these ultra-large scales, primordial non-Gaussianity leaves signatures that can shed light on the 
mechanism by which perturbations in the early Universe are generated. 
We perform a forecast analysis for constraining  local type non-Gaussianity and its two-parameter 
extension with a simple scale-dependence. 
We combine different clustering measurements from future galaxy surveys -- a 21cm intensity mapping survey and two photometric galaxy 
surveys -- via the multi-tracer approach. 
Furthermore we then include CMB lensing from a CMB Stage 4 experiment in the multi-tracer, which can 
improve the constraints on bias parameters.
We forecast $\sigma(f_{\rm NL}) \simeq 0.9$ (1.4) by combining SKA1, a Euclid-like (LSST-like) survey, 
and CMB-S4 lensing.
With CMB lensing, the precision on $f_{\rm NL}$ improves by up to a factor of 2, showing that a joint analysis 
is important. 
In the case with running of $f_{\rm NL}$, our results show that the combination of upcoming 
cosmological surveys could achieve $\sigma(n_{\rm NL}) \simeq 0.12$ (0.22) on the running index.
\end{abstract}




\section{Introduction}
\label{sec:intro}

The coherent nature of the cosmic microwave background (CMB) anisotropies and the 
large-scale structure (LSS) we observe around us suggests that the seed for these 
fluctuations were created at very early times, possibly during a period of inflation 
\citep{Starobinsky:1980te,Guth:1980zm,Sato:1980yn,Linde:1981mu,Albrecht:1982wi,Hawking:1982ga,Linde:1983gd,Mukhanov:1981xt}.

Inflation observables are predicted to be proportional to the slow-roll parameters for 
the single field slow-roll (SFSR) models and to be connected through consistency 
relations for this simplest class of models. For this reason, in the absence of any 
salient features in the primordial power spectrum, which might open a new observational 
window on high energy physics happening in the early Universe \citep{Chen:2014cwa}, 
SFSR constraints in the next decade will likely be limited to improvements to the constraints 
on the scalar spectral index and the tensor-to-scalar ratio.
The prospects of detecting the running of the scalar spectral index, that 
arises in SFSR models at second-order in slow-roll parameters ($\dd n_s/\dd \ln k \propto (n_s-1)^2$), 
may be nearly impossible even with 
next-generation cosmological surveys \citep{Ballardini:2016hpi,Munoz:2016owz,Li:2018epc,Mifsud:2019xpu}.

One additional observational probe that allows us access to early-Universe physics  
is primordial non-Gaussianity (PNG) \citep[see][for reviews]{Bartolo:2004if,Chen:2010xka}.
The PNG parameter $f_{\rm NL}$ is predicted to be first
order in slow-roll from consistency relations for SFSR model, $f_{\rm NL} \simeq -5(n_s-1)/12$ 
\citep{Acquaviva:2002ud,Maldacena:2002vr,Creminelli:2004yq}.
On the other hand, an $f_{\rm NL}\gtrsim 1$ is expected for many multi-field inflation models 
\citep[see][for a review]{Byrnes:2010em}.

At present, the best constraints on PNG come from {\em Planck} measurements 
of the three-point correlation function of the CMB temperature and polarization 
anisotropies \citep{Akrami:2019izv}, but LSS is emerging as 
a promising complementary observable.
Nonlinear mode coupling from local PNG induces a modulation of 
the local short-scale power spectrum through a scale dependence in the bias produced 
by the long-wavelength primordial gravitational potential 
$\Phi$ \citep{Salopek:1990jq,Gangui:1993tt}
\begin{equation} \label{eqn:fNL}
    \Phi({\bf x}) = \phi({\bf x}) + f_{\rm NL}\left(\phi^2({\bf x})-\langle\phi^2\rangle\right) 
                    + {\cal O}\left(\phi^3\right)
\end{equation}
where $\phi$ is a Gaussian field.
The appearance of $\Phi$ in the halo bias implies a specific form of scale-dependence 
that cannot be created dynamically (i.e. by late time processes).  

This is the main reason that halo bias is such a robust probe of the initial conditions 
and this gives us the opportunity to study PNG with two-point statistics of the LSS.
Crucially, the $k^{-2}$ scaling which arises for some {\it local} model of PNG 
makes the signal largest on the very largest scales of the matter 
power spectrum \citep{Dalal:2007cu,Matarrese:2008nc,Desjacques:2008vf,Slosar:2008hx,Camera:2013kpa}.
Such large scales, greater than the equality scale, are affected strongly by cosmic variance,
which puts a fundamental limit on the precision with which $f_{\rm NL}$ can be measured \citep{Alonso:2015uua}.

A novel proposal to improve the expected constraints on the amplitude of the PNG 
fluctuations is to combine the information coming from different 
LSS tracers or to split the sample in bins of different halo mass in order to 
reduce the sample variance \citep{Seljak:2008xr,Yoo:2012se,Abramo:2013awa,Ferramacho:2014pua,Yamauchi:2014ioa,Ferraro:2014jba,dePutter:2014lna,Alonso:2015sfa,Fonseca:2015laa,Fonseca:2016xvi,Abramo:2017xnp,Fonseca:2018hsu}. This is the so-called {\it multi-tracer} 
approach.
Moreover, the cross-correlation between clustering and CMB lensing has recently been 
shown to be particularly well-suited to measure local PNG using the 
scale-dependent halo bias \citep{Schmittfull:2017ffw,Giusarma:2018jei}.
The cross-correlation between CMB lensing and clustering has a high 
signal to noise and decreases the total effective variance compared to the case considering 
the two fields independently.

Currently, the tightest constraints on local
type PNG are $f_{\rm NL} = -0.9 \pm 5.1$ at  68\% CL from the {\em  Planck} 2018 
data \citep{Akrami:2019izv}, and $-51 < f_{\rm NL} < 21$ at 95\% CL from eBOSS DR14 data \citep{Castorina:2019wmr}.

This paper aims to assess the constraining power achievable by a multi-tracer combination 
of two next-generation galaxy surveys and a CMB Stage 4 (CMB-S4) survey.

We consider also a generalization of the $f_{\rm NL}$-model \eqref{eqn:fNL}, in which the parameter 
$f_{\rm NL}$ is promoted to a function of scale $k$ \citep{Chen:2005fe,Byrnes:2008zy,Byrnes:2009pe,Raccanelli:2014awa}
\begin{equation} \label{eqn:nNL}
    f_{\rm NL}(k) = f_{\rm NL} \left(\frac{k}{k_{\rm piv}}\right)^{n_{\rm NL}},
\end{equation}
where $k_{\rm piv}$ is some pivot scale fixed at 0.035 h/Mpc.
The tightest current observational constraint on the running index is
from the bispectra of the CMB fluctuations: 
$-0.6< n_{\rm NL}<1.4$ at  68\% CL from WMAP9 data, 
for the single-field curvaton scenario \citep{LoVerde:2007ri,Sefusatti:2009xu,Becker:2012je,Oppizzi:2017nfy}.

This paper is organized as follows: in section~\ref{sec:two} we describe how the different 
PNG templates enter into the halo bias through a scale-dependent contribution.
We then describe the cosmological surveys considered in our analysis: CMB-S4 
as a CMB experiment, SKA1-MID Band 1 IM, and as LSS experiments: Euclid-like and LSST-like.
in section~\ref{sec:three}. We also introduce the Fisher forecasting formalism in section~\ref{sec:three}.
Finally, we present our results in section~\ref{sec:four} and we draw our conclusion in 
section~\ref{sec:conc}.

\section{Primordial non-Gaussianity and Large Scale Structure}
\label{sec:two}

In this section we describe the large-scale halo bias in the context of the 
peak-background split (PBS) \citep{Mo:1995cs,Sheth:1999mn,Schmidt:2010gw,Desjacques:2016bnm}. 
The PBS method is used to predict the large-scale clustering statistics of 
dark matter halos. The Gaussian field is split into 
long- and short-scale modes $\phi = \phi_\ell + \phi_s$, where the long scales determine 
the clustering of halos relevant for large-scale power spectrum analysis, 
while the short scales govern the halo formation.

In order to connect the comoving matter density contrast $\delta$ to the gravitational 
potential $\Phi$, we make use of the Poisson equation at late times
\begin{equation} \label{eqn:PoissLate}
    \nabla ^2 \Phi = -\frac{3}{2}\Omega_{m,0}H_0^2\frac{\delta}{a}
\end{equation}
where  the potential has been defined under the following convention for the perturbed metric in the Newtonian gauge
\begin{equation}
    a^{-2} \dd s^2 =  (1+2\Psi) \dd\eta^2-(1-2\Phi) \dd x_i \dd x^i \,.
\end{equation}
At late times the gravitational potential $\Phi$ can be connected to the primordial potential 
$\Phi_{\rm p}$ by 
\begin{equation}
    \Phi({\bf k},z) = \frac{T(k) \Phi_{\rm p}({\bf k})D(z)}{a(z)}
\end{equation}
where $T(k)$ is the matter transfer function normalized to one at ultra-large scales, 
and $D(z)$ is the growth factor normalized to the scale factor in the matter-dominated era.

From the Poisson Eq.~\eqref{eqn:PoissLate}, we can write the matter density contrast as 
\begin{equation} \label{eqn:deltaG}
    \delta({\bf k},z) = \alpha(k,z) \Phi_{\rm p}({\bf k}) 
\end{equation}
with 
\begin{equation}
    \alpha(k,z) \equiv \frac{2 k^2 T(k) D(z)}{3 \Omega_{m,0}H_0^2} \,.
\end{equation}

In the presence of local PNG of the form Eq.~\eqref{eqn:fNL}, the Laplacian of the primordial 
potential is
\begin{equation}
    \nabla^2 {\Phi_{\rm p}} \simeq \nabla^2 \phi + 2 f_{\rm NL} \left(\phi\nabla^2\phi+|\nabla\phi|^2\right)
\end{equation}
and we can split its contribution into long and short wavelengths at leading order as 
\begin{align}
    &\Phi_{\rm l} \approx \phi_{\rm l} ,\\
    &\Phi_{\rm s} \approx \phi_{\rm s} \left(1+2f_{\rm NL} \phi_{\rm l}\right) \label{eqn:Phis} \,.
\end{align}

The long-wavelength overdensity $\delta_{\rm l}$ which describes the clustering properties 
of the matter distribution is not affected by the presence of PNG 
\begin{equation}
    \delta_{\rm l}({\bf k},z) = \alpha(k,z) \phi_{\rm l}({\bf k}) \,,
\end{equation}
while the short-wavelength fluctuations are altered by long wavelengths. At lowest order,
neglecting white-noise contributions, we have
\begin{equation}
    \delta_{\rm s}({\bf k}) = \alpha(k,z) \phi_{\rm s}({\bf k})\left(1 + 2f_{\rm NL} \phi_{\rm l}\right) \,.
\end{equation}

The local number density of halos in Lagrangian space is given by
\begin{equation}
    n_{\rm h} = \bar{n}_{\rm h} \left(1 + b_{\rm L} \delta_{\rm l} \right)
\end{equation}
where $b_{\rm L}$ is the Lagrangian-space bias and $\delta_{\rm l}$ is again the contribution 
from the long-wavelength modes in \eqref{eqn:Phis} that essentially modulate the mean density 
of the effective local cosmology. Therefore
\begin{equation}
    b_{\rm L} = \frac{\dd \ln n_{\rm h}}{\dd \delta_{\rm l}}\,,
\end{equation}
and the more usual Eulerian-space bias is given by $b=1 +b_{\rm L}$.

In the presence of PNG, the local number of halos does not just depend on the large-scale 
matter perturbations, but it is also affected by the mode coupling between long 
and short wavelengths that acts like a local rescaling of the amplitude of (small-scale) matter
fluctuations. Taylor expanding at first order in these parameters
\begin{align}
    b_{\rm L} &= \frac{\dd \ln n_{\rm h}}{\dd \delta_{\rm l}} \notag\\
    &= \frac{\partial \ln n_{\rm h}}{\partial \delta_{\rm l}} 
     + \frac{\partial \left(1+2f_{\rm NL}\phi_{\rm l}\right)}{\partial \phi_{\rm l}}\frac{\partial \phi_{\rm l}}{\partial \delta_{\rm l}}\frac{\partial \ln n_{\rm h}}{\partial \left(1+2f_{\rm NL}\phi_{\rm l}\right)} \notag\\
    &= \frac{\partial \ln n_{\rm h}}{\partial \delta_{\rm l}} + \frac{2 f_{\rm NL}}{\alpha(k,z)}\frac{\partial \ln n_{\rm h}}{\partial \ln \sigma_8^{\rm loc}} \\
    &= b_{\rm L}^{\rm Gauss} + \Delta b(k,z)
\end{align}
where we parametrize the local amplitude of small-scale fluctuations by  $\sigma_8^{\rm loc} = \sigma_8\left(1+2f_{\rm NL}\phi_{\rm l}\right)$, and we 
introduce the scale-dependent contribution to the large-scale bias as 
\begin{equation}
    \Delta b(k,z) = f_{\rm NL}\frac{\beta_f}{\alpha(k,z)} \,.
\end{equation}

Finally, on large scales we can relate the halo density contrast  to the linear 
density field as
\begin{equation}
    \delta_{\rm h}({\bf k},z) = \left[b(z) + \Delta b(k,z)\right]\delta({\bf k},z) 
\end{equation}
where $b = 1 + b_{\rm L}^{\rm Gauss}$ is the Eulerian-space bias connected to the Gaussian 
Lagrangian-space bias.

Throughout this paper, we will use the expression $\beta_f = 2 \delta_{{\rm c}}(b-1)$, 
which is exact in a barrier crossing model with barrier height $\delta_{\rm c}$ and is a good 
($\approx 10\%$ accuracy) fit to N-body simulations \citep{Dalal:2007cu,Biagetti:2016ywx}. 
We see that, unlike the Gaussian 
linear bias $b$, the non-Gaussian linear bias 
will no longer be scale-independent, correcting $b$ by a factor $\propto f_{\rm NL}/k^2$.

Note that there are two conventions to define $f_{\rm NL}$ in Eq.~\eqref{eqn:fNL}: the LSS convention, 
where $\Phi$ is normalized at $z=0$, so that $D(0)=1$, and the CMB convention where $\Phi$ is instead the primordial 
potential, so that $D(a)=a$ in the matter dominated era. The relation between the two normalizations is
\begin{equation} \label{eqn:convention}
    f_{\rm NL}^{\rm LSS} = \frac{D(z=\infty)(1+z)}{D(z=0)} f_{\rm NL}^{\rm CMB} \,.
\end{equation}
We adopt the CMB convention.

\section{Setup}
\label{sec:three}

We describe in this section the specifications for the different cosmological surveys 
used in the analysis and the details of the Fisher methodology used to infer uncertainties 
on $f_{\rm NL}$ and $n_{\rm NL}$.

\subsection{CMB lensing specifications}

We work with a possible CMB-S4 configuration assuming a 3 arcmin beam and 
$\sigma_{\rm T}=\sigma_{\rm P}/\sqrt{2}=1\,\mu$K-arcmin noise \citep{Abazajian:2016yjj}. 
We assume $\ell_{\rm min}=30$ and a different cut at high-$\ell$ of $\ell_{\rm max}^{\rm T}=3000$ 
in temperature and $\ell_{\rm max}^{\rm P}=5000$ in polarization, with $f_{\rm sky} = 0.4$.

For CMB temperature and polarization  angular power spectra, the instrumental noise 
deconvolved with the instrumental beam is defined by \citep{Knox:1995dq}
\begin{equation}
    {\cal N}_\ell^{\rm T,P} = \sigma_{\rm T,P} b_\ell^{-2}\,,
\end{equation}
where we assume a Gaussian beam
\begin{equation} \label{eqn:beam}
    b_\ell = \exp\left[-\ell(\ell+1)\frac{\theta^2_{\rm FWHM}}{16 \ln 2}\right] \,.
\end{equation}

For CMB lensing, we assume that the lensing reconstruction can be performed with the minimum 
variance quadratic estimator on the full sky, combining the TT, EE, BB, TE, TB, and EB  estimators, 
calculated according to \cite{Hu:2001kj} with {\tt quicklens}\footnote{\href{https://github.com/dhanson/quicklens}{https://github.com/dhanson/quicklens}} and 
applying iterative lensing reconstruction \citep{Hirata:2003ka,Smith:2010gu}.
We use the CMB lensing information in the range $30 \leq \ell \leq 3000$.

Note that hereafter we will refer to the full set of angular power spectra of the CMB anisotropies
(i.e. temperature, E-mode polarization, CMB lensing, and their cross-correlations) as simply 
`CMB'.

\subsection{HI intensity mapping specifications}

IM surveys measure the total intensity emission in each pixel for given atomic lines with very accurate redshifts, without resolving individual galaxies, which are hosts of the emitting atoms \citep{Battye:2004re,Wyithe:2007gz,Chang:2007xk,Bull:2014rha,Santos:2015bsa,Kovetz:2017agg}.
The measured brightness temperature 
fluctuations are expected to be a biased tracer of the underlying cold dark matter distribution.

We consider neutral hydrogen (HI) 21cm emission and we use the fitting formulas from \citet{Santos:2017qgq} for the 
HI linear bias:
\begin{equation}
    b_{\rm HI}(z) = \frac{b_{{\rm HI}}(0)}{0.677105} \Big[0.66655 + 0.17765\, z + 0.050223\,  z^2\Big],  
\end{equation}
and for the background HI brightness temperature:
\begin{equation}
    \bar{T}_{{\rm HI}}(z) = 0.055919 + 0.23242\, z - 0.024136\, z^2\ \text{mK},
\end{equation}
where $\Omega_{\rm HI}(0)b_{\rm HI}(0) = 4.3\times 10^{-4}$ and 
$\Omega_{\rm HI}(0) = 4.86\times 10^{-4}$.

The noise variance for IM with $N_{\rm dish}$ dishes in single-dish mode in the frequency $i$-channel, assuming scale-independence 
and no correlation between the noise in different frequency channels, is \citep{Knox:1995dq,Bull:2014rha}
\begin{align} \label{eqn:Nhi}
    \sigma_{\rm HI}({\nu_i}) &= \frac{4\pi f_{\rm sky}T^2_{\rm sys}({\nu_i})}{2N_{\rm dish}t_{\rm tot}\Delta\nu} \,,\\
    T_{\rm sys}({\nu_i}) &= 25 + 60\left(\frac{300\,\text{MHz}}{{\nu_i}}\right)^{2.55}\ \text{K} \,,
\end{align}
where $t_{\rm tot}$ is the total observing time.
We also include the instrumental limit in angular resolution, characterized by the telescope beam. 
We assume the noise deconvolved with a Gaussian beam modelled as
\begin{equation}
    {{\cal N}_\ell^{\rm HI}}({\nu_i}) = \sigma_{\rm HI}b_\ell^{-2}({\nu_i})\,,
\end{equation}
where $b_\ell({\nu_i})$ is the contribution of the beam in the 
frequency $i$-channel given by Eq.~\eqref{eqn:beam} with
\be
\theta_{\rm FWHM}(\nu) \approx \frac{c}{\nu D_{\rm dish}} \,.
\ee

For SKA1-MID, we assume $N_{\rm dish} = 197$, $D_{\rm dish} = 15$ m, $t_{\rm tot} = 10^4$ 
hr observing over 20,000\,deg$^2$ in the redshift range $0.35 \le z \le 3.05$ 
($1050 \geq \nu \geq 350\,$MHz, Band 1) \citep{Bacon:2018dui}. We divide  the redshift range into 27 tomographic bins 
with  width 0.1. The cleaning of foregrounds from HI IM effectively removes the largest scales, $\ell_{\rm min} \lesssim 5$ 
\citep{Witzemann:2018cdx,Cunnington:2019lvb}, and we take $\ell_{\rm min}=5$.

\subsection{Galaxy survey specifications}

We present the details of two future photometric galaxy surveys. For each survey we assume 
the  redshift distribution of sources of the form
\begin{equation} \label{eqn:dNdz}
    {n_g}(z) \propto z^{\alpha} \exp \left[ -\left( \frac{z}{z_0} \right)^{\beta} \right] \mbox{gal/arcmin}^2\,.
\end{equation}
The distribution of sources in the $i$-th redshift bin, including 
photometric uncertainties, following \citep{Ma:2005rc}, is
\begin{equation}
    {n^i_g}(z) = \int_{z^i_{\rm ph}}^{z^{i+1}_{\rm ph}} \dd z_{\rm ph}\, {n_g}(z)\, p(z_{\rm ph}|z)\,,
\end{equation}
where we adopt a Gaussian distribution for the probability distribution of photometric redshift 
estimates $z_{\rm ph}$, given true redshifts $z$:
\begin{equation}
    p(z_{\rm ph}|z) = \frac{1}{\sqrt{2\pi}\,\sigma_z} \exp \Bigg[-\frac{\big(z-z_{\rm ph}\big)^2}{2\sigma_z^2}\Bigg] \,.
\end{equation}

The shot noise for galaxies in the $i$-th redshift bin is the inverse of the angular number density of galaxies: 
\begin{equation}
   {{\cal N}^{gi}_\ell = 
    \left( \int \dd z\ {n^i_g}(z)\right)^{-1}}\,.
\end{equation}

Finally, we impose a cut on small scales assuming that we will be able to reconstruct 
non-linear scales up to $k_{\rm max} = 0.3\ h/$Mpc, which corresponds to a redshift-dependent 
cut in angular space:  {$\ell_{\rm max} \simeq \chi(z)k_{\rm max}$}.

\subsubsection{Euclid-like survey}

The Euclid satellite is a mission of the ESA Cosmic Vision program that will be launched in 2022 
\citep{Laureijs:2011gra}. It will perform both a photometric and spectroscopic survey of galaxies. 
In this work, we focus only on a Euclid-like photometric survey that will cover 
${\Omega}_{\rm sky} = 15,000\,$deg$^2$ measuring ${n_g}=30\,$sources per arcmin$^2$ over 
a redshift range $0<z<2.5$ \citep{Amendola:2016saw}.

The redshift distribution  follows Eq.~\eqref{eqn:dNdz},
with $\alpha$ = 2, $\beta$ = 1.5, and $z_{0}$ = 0.636, divided into 10 bins each containing the 
same number of galaxies \citep{Amendola:2016saw}.
The scatter of the photometric redshift estimate with respect to the true redshift value is 
$\sigma_z = 0.05 (1+z)$. The fiducial model for the  linear bias is 
$b(z)= \sqrt{1+z}$ \citep{Amendola:2016saw}. We assume ${\ell}_{\rm min}=10$.

\subsubsection{LSST-like survey}

For LSST clustering measurements, we assume a number density of galaxies of ${n_g}=48$ 
sources per arcmin$^2$ observed over a patch  ${\Omega}_{\rm sky} = 13,800$ deg$^2$ and 
distributed in redshift according to Eq.~\eqref{eqn:dNdz}, with $\alpha$ = 2, $\beta$ = 0.9, and 
$z_{0}$ = 0.28 \citep{Mandelbaum:2018ouv}.

We assume 10 tomographic bins spaced by $0.1$  between $0.2 \leq z \leq 1.2$, with 
photometric redshift uncertainties $\sigma_z = 0.03 (1+z)$, and a fiducial model for the   bias given by 
$b(z)= 0.95/D(z)$
\citep{Mandelbaum:2018ouv}. We impose ${\ell}_{\rm min}=20$.

\subsection{Fisher analysis}

We use the Fisher matrix to derive forecasted constraints on the cosmological parameters, 
assuming that the observed fields are Gaussian random distributed (for simplicity we 
ignore information from higher-order statistics).

The Fisher matrix at the power spectrum level is then
\begin{equation} \label{eqn:fisher}
    F_{ij} = f_{\rm sky} \sum_{\ell=\ell_{\min}}^{\ell_{\rm max}} \left(\frac{2\ell+1}{2}\right)\,
    \text{tr} \big[{\bf C}_{\ell,i}\,{\text{\boldmath$\Sigma$}}_\ell \, {\bf C}_{\ell,j}\,{\text{\boldmath$\Sigma$}}_\ell\big]\,,
\end{equation}
where ${\bf C}_{\ell}$ is the covariance matrix, 
${\bf C}_{\ell,i}=\partial {\bf C}_{\ell}/\partial \theta_i$ is the derivative 
with respect to the $i$-th  cosmological parameter, and 
${\text{\boldmath$\Sigma$}}_{\ell}=\left({\bf C}_{\ell}+{\bf N}_{\ell}\right)^{-1}$ is the inverse of 
the total noise matrix, with ${\bf N}_{\ell}$ the diagonal noise matrix.
This equation assumes that all experiments observe the same patch of sky. 
We consider for each experiment its own sky fraction and for the cross-correlations the smallest of the sky 
fractions.

The angular power spectra are 
\begin{equation} \label{eqn:APS}
    C_{\ell}^{XY}{(z_i,z_j)} = 4 \pi \int \frac{\dd k}{k}\, {\cal P_R} (k)\, I_\ell^{X} (k,{z_i}) \, I_\ell^{Y} (k,{z_j}) \,. 
\end{equation}
Here  $X,Y = {\rm T}, {\rm E}, \phi$ for the CMB, and $=\Delta_g,\Delta_{\rm HI}$ for the galaxy clustering/ IM surveys,  where $\Delta_g=\delta_g+\,$observational corrections from observing on the past lightcone, and similarly for $\Delta_{\rm HI}$ \citep[see][for details]{Challinor:2011bk,Ballardini:2018cho}.
 ${\cal P_R}$ is the dimensionless primordial power spectrum and the large-scale structure kernels are
\begin{align}
    I_\ell^{\Delta_g} (k,z_{i}) &= \int \dd z\ {n^i_g(z) \Delta^g_\ell(k,z)}\,,\\
    I_\ell^{\Delta_{\rm HI}} (k,z_{i}) &= \int \dd z\  W_{\rm th}(z,z_i) \bar{T}_{\rm HI}(z) {\Delta^{\rm HI}_\ell(k,z)}\,,  
\end{align}
where $\Delta^g_\ell, \Delta^{\rm HI}_\ell$ are the angular transfer functions \citep[see][]{Ballardini:2018cho} and $W_{\rm th}(z,z_i)$ is a smoothed top-hat window function for the $i$-th bin.
We refer the reader to \citet{Hu:1997hp} for the details of the CMB temperature and polarization 
window functions.

All the angular power spectra have been calculated using a modified version of the publicly 
available code {\tt CAMB} \footnote{\href{https://github.com/cmbant/CAMB}{https://github.com/cmbant/CAMB}} 
\citep{Lewis:1999bs,Howlett:2012mh,Challinor:2011bk}.

\section{Results}
\label{sec:four}

The standard cosmological parameter vector that we use is 
\begin{equation}
    \text{\boldmath$\theta$} = \left\{\omega_b,\omega_c,H_0,\tau,\ln \left(10^{10}A_s\right),n_s\right\} .
\end{equation}
In addition, we have the PNG parameters depending on the model studied: $\left\{f_{\rm NL}\right\}$ or 
$\left\{f_{\rm NL}, n_{\rm NL}\right\}$.
We also include a nuisance parameter for each redshift bin, in each of the LSS surveys,  allowing 
for a free redshift evolution of the clustering bias ${b}$ or of the combination 
${\bar{T}_{\rm HI} b_{\rm HI}}$ for IM.

The fiducial cosmology used for the standard cosmological parameters, according to 
{\em Planck} 2018 \citep{Aghanim:2018eyx}, is: 
$\omega_b = 0.022383$, 
$\omega_c = 0.12011$, 
$H_0 = 67.32$, 
$\tau = 0.0543$, 
$\ln \left(10^{10}A_s\right) = 3.0448$, 
$n_s = 0.96605$.
We assume as fiducial $f_{\rm NL} = 0$ without running and $f_{\rm NL} = -0.9$, $n_{\rm NL} = 0$ 
for the extended model.

Uncertainties reported in the following subsections have been marginalized over 
all the 6 standard cosmological parameters and the nuisance bias parameters, 
i.e. 27 temperature-bias parameters for SKA1-HI IM and 10 galaxy bias parameters 
for Euclid-like/ LSST-like.

\subsection{$f_{\rm NL}$ model of PNG}

\begin{figure}
\includegraphics[width=.5\textwidth]{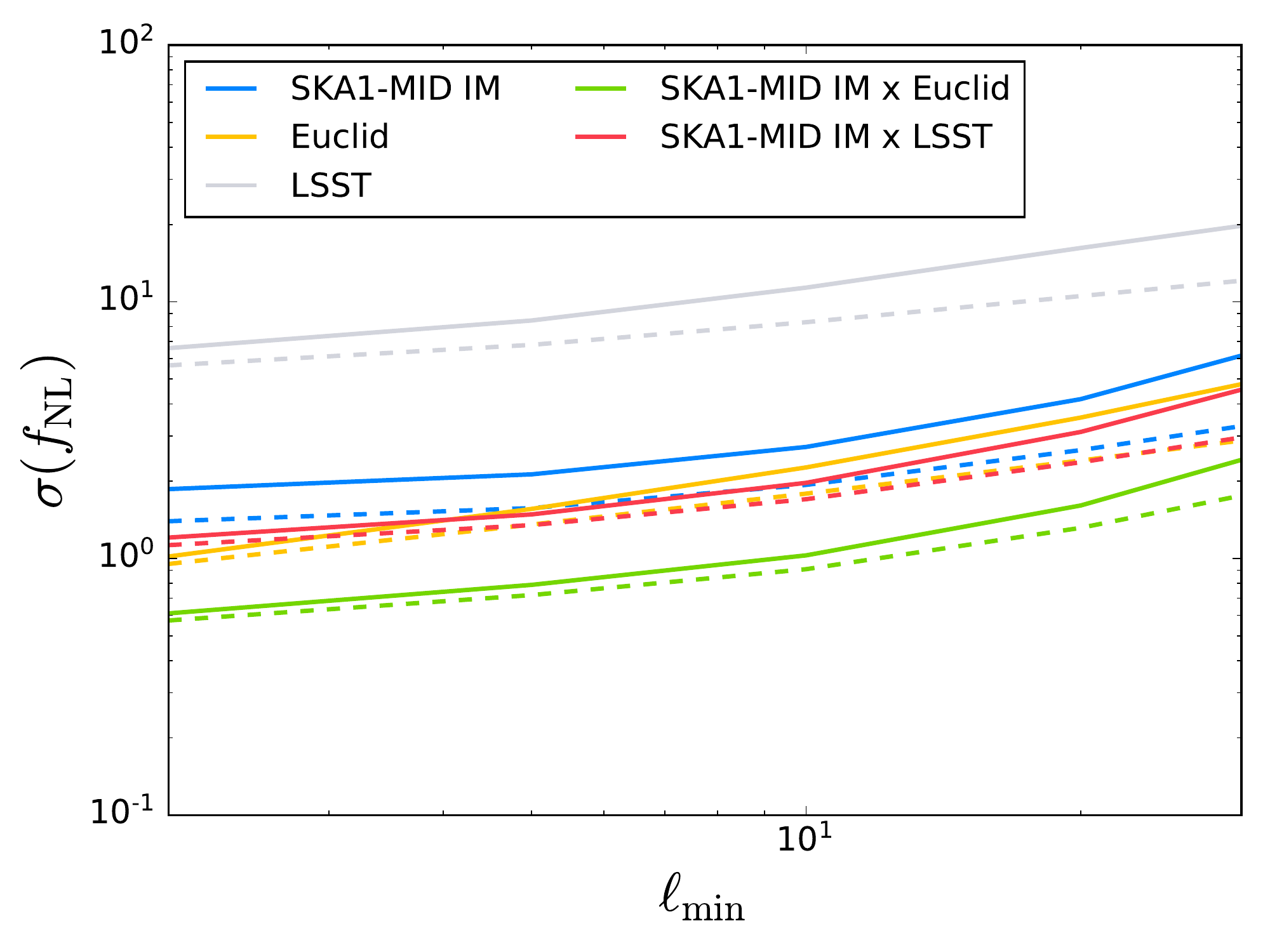}
\caption{Marginalized uncertainties on $f_{\rm NL}$ as function of the minimum multipole 
$\ell_{\rm min}$ of LSS. Solid curves correspond to LSS experiments 
without CMB: SKA1 IM (blue), Euclid-like (yellow), LSST-like (grey), and the combinations SKA1 IM $\times$ Euclid-like (green) and LSST-like (red). 
Dashed lines correspond to 
the inclusion of CMB-S4 lensing ($\ell_{\rm min}=30$).}
\label{fig:fNL_lmin}
\end{figure}

We consider different minimum multipoles as feasible for the different experimental configurations 
described in section~\ref{sec:three}. In figure~\ref{fig:fNL_lmin}, we present the uncertainties 
on $f_{\rm NL}$ as a function of $\ell_{\rm min}$.

The uncertainties for single surveys, with the assumed minimum multipole, are:
\begin{equation}\label{sf1}
\sigma\left( f_{\rm  NL} \right) \simeq \begin{cases} 2.1 & \mbox{SKA1 }(\ell_{\rm min}=5)\,, \\ 2.3 & \mbox{Euclid-like } (\ell_{\rm min}=10)\,, \\
16.2 & \mbox{LSST-like } (\ell_{\rm min}=20)\,.
\end{cases}
\end{equation}

Including CMB lensing from CMB-S4 with $\ell_{\rm min}=30$,
using the above $\ell_{\rm min}$ values for LSS and the smallest sky area as the overlap area, the  errors in Eq. \eqref{sf1} decrease to:
\begin{equation}\label{sf2}
\sigma\left( f_{\rm  NL} \right) \simeq \begin{cases} 1.6 & \mbox{SKA1}\times \mbox{CMB-S4}\,, \\ 1.8 & \mbox{Euclid-like}\times \mbox{CMB-S4} \,, \\
10.5 & \mbox{LSST-like}\times \mbox{CMB-S4} \,.
\end{cases}
\end{equation}

The combination of intensity and number counts, using the above $\ell_{\rm min}$ values and the smaller sky area as the overlap area,
leads to the errors:
\begin{equation}\label{sf3}
\sigma\left( f_{\rm  NL} \right) \simeq \begin{cases} 0.96 & \mbox{SKA1}\times \mbox{Euclid-like}\,, \\ 1.6 & \mbox{SKA1}\times \mbox{LSST-like} \,.
\end{cases}
\end{equation}

When  all three tracers are combined, the tightest constraints obtained are
\begin{equation}\label{sf4}
\sigma\left( f_{\rm  NL} \right) \simeq \begin{cases} 0.90 &  \mbox{SKA1}\times \mbox{Euclid-like} \times \mbox{CMB-S4}\,,\\
1.4 &  \mbox{SKA1}\times \mbox{LSST-like} \times \mbox{CMB-S4}\,.
\end{cases}
\end{equation}


In addition, we investigate the optimistic case where the minimum multipoles extend down to $\ell_{\rm min}=2$ for all three tracers. 
This yields the following constraints for the full multi-tracer cases: 
\begin{equation}\label{sf5}
\sigma\left( f_{\rm  NL} \right) \simeq \begin{cases} 0.47 &  \mbox{SKA1}\times \mbox{Euclid-like} \times \mbox{CMB-S4}\,,\\
1.0 &  \mbox{SKA1}\times \mbox{LSST-like} \times \mbox{CMB-S4}\,.
\end{cases}
\end{equation}

\subsection{$f_{\rm NL}, n_{\rm NL}$ model of PNG}

\begin{figure}
\includegraphics[width=.5\textwidth]{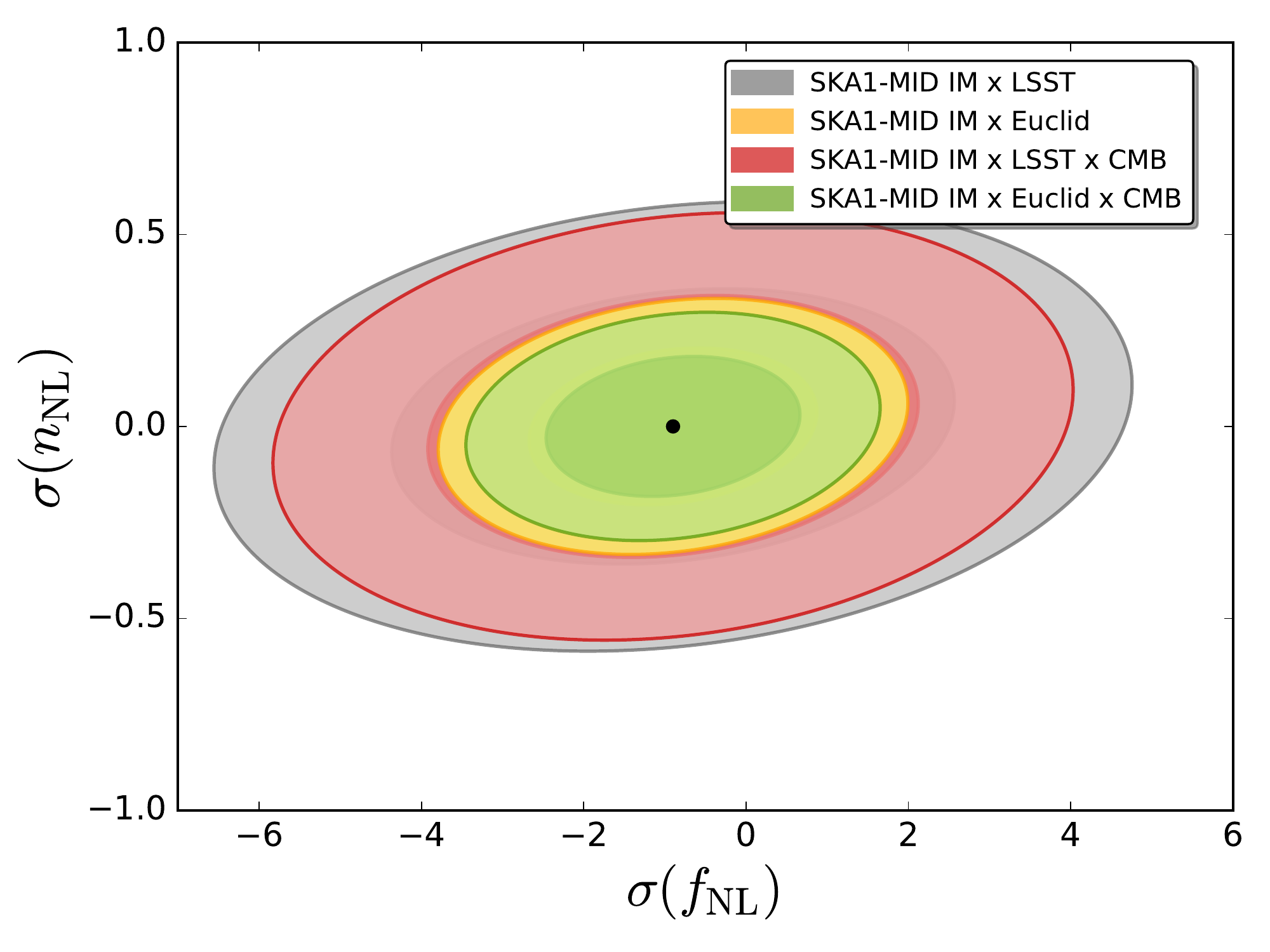}
\caption{Marginalized 2-dimensional contours (68\% and 95\% CL) for $f_{\rm NL}$ and $n_{\rm NL}$, with 
$\ell_{\rm min} =$ 5, 10, 30 for HI IM, galaxy number counts, and CMB respectively. 
The multi-tracer combinations are: SKA1 $\times$ Euclid-like (yellow), 
SKA1 $\times$ LSST-like (grey), CMB $\times$ SKA1 $\times$ Euclid-like (green), 
and CMB $\times$ SKA1 $\times$ LSST-like (red).}
\label{fig:nNL-2D}
\end{figure}

We turn now to the constraints for the two-parameter model \eqref{eqn:nNL} with a running of $f_{\rm NL}$, using the same specification as in Eqs. \eqref{sf1}--\eqref{sf4}.
Figure~\ref{fig:nNL-2D} shows the marginalized uncertainties on the 2-dimensional 
$(f_{\rm NL},n_{\rm NL})$ parameter space.

The uncertainties for single tracers are 
\begin{equation}\label{sn1}
\sigma\left( n_{\rm  NL} \right) \simeq \begin{cases} 2.7 & \mbox{SKA1 }(\ell_{\rm min}=5)\,, \\ 0.35 & \mbox{Euclid-like } (\ell_{\rm min}=10)\,, \\
0.37 & \mbox{LSST-like } (\ell_{\rm min}=20)\,.
\end{cases}
\end{equation}

Including CMB lensing from CMB-S4 with $\ell_{\rm min}=30$, errors decrease to 
\begin{equation}\label{sn2}
\sigma\left( n_{\rm  NL} \right) \simeq \begin{cases} 1.4 & \mbox{SKA1}\times \mbox{CMB-S4}\,, \\ 0.24 & \mbox{Euclid-like}\times \mbox{CMB-S4} \,, \\
0.32 & \mbox{LSST-like}\times \mbox{CMB-S4} \,.
\end{cases}
\end{equation}

The combination of intensity and number counts 
leads to 
\begin{equation}\label{sn3}
\sigma\left( n_{\rm  NL} \right) \simeq \begin{cases} 0.13 & \mbox{SKA1}\times \mbox{Euclid-like}\,, \\ 0.24 & \mbox{SKA1}\times \mbox{LSST-like} \,.
\end{cases}
\end{equation}

When all three tracers are combined, the tightest constraint obtained is
\begin{equation}\label{sn4}
\sigma\left( n_{\rm  NL} \right) \simeq \begin{cases} 0.12 &  \mbox{SKA1}\times \mbox{Euclid-like} \times \mbox{CMB-S4}\,,\\
0.22 &  \mbox{SKA1}\times \mbox{LSST-like} \times \mbox{CMB-S4}\,.
\end{cases}
\end{equation}

In this case the uncertainties on $f_{\rm NL}$ degrade by $\sim20\%$ on average, compared 
to the case without running, which shows a weak degeneracy between the two parameters.

\subsection{Comparison with other results on $\sigma\left( f_{\rm  NL} \right)$}

In this work, we consistently make use of the CMB convention to define $f_{\rm NL}$. 
In comparison with other work where the alternative LSS convention is used, we quote here the relevant constraints modified to be consistent with the CMB convention \eqref{eqn:convention}.

In \citet{Alonso:2015uua} and \citet{Alonso:2015sfa}, the case of LSST-like and SKA1-MID IM is treated (without using CMB-S4), 
giving uncertainties down to $\sim 0.31$ for the multi-tracer case. They use a greater number of thinner bins for the SKAI-MID IM survey, i.e., 100 bins with equal co-moving width, while we use 27 such bins. For the LSST-like survey, they use 9 bins 
with widths chosen to ensure equal source density, as opposed to our 10 
fixed-width bins. 
They use  a multipole range $2\leq \ell \leq 500$ for both tracers, and assume larger sky fractions:
0.5 for LSST-like and 0.75 for SKA1, with the overlap taken as 0.4, which also exceeds ours.
Their LSST-like redshift distribution has a slightly more pessimistic 40 sources/arcmin$^2$, versus our 
48 sources/arcmin$^2$ according to \citet{Mandelbaum:2018ouv}, which results in a slightly lower shot noise. In summary, their greater sky area and smaller $\ell_{\rm min}$ are the main reasons for their more optimistic constraints.

In \citet{Fonseca:2015laa} there is a multi-tracer analysis for Euclid-like and SKA1-MID IM surveys. Their  results give $0.72\leq \sigma\left( f_{\rm NL} \right)\leq 1.05$,  depending on (a)~the maximum multipole chosen ($\ell_{\rm max} = 60$ 
or $\ell_{\rm max} = 300$), and (b)~the sky overlap (50\% or 100\%).  Their multipole range for all tracers extends down to $\ell_{\rm min} = 2$.
They also consider a LSST-like survey with sky area equal to that of the entire 
SKA1-MID IM. They obtain the multi-tracer result $\sigma\left( f_{\rm NL} \right) \simeq 0.61$ for $\ell_{\rm max} = 300$, which is lower than ours.
 Considering that the effect of $f_{\rm NL}$ is captured only on larger scales, this difference 
in $\ell_{\rm max}$ should have a negligible effect on the final uncertainties. 
The sky fraction in their 50\% overlap case is 0.18, which is smaller than our 
shared sky fraction of 0.36 for SKA1-MID IM and Euclid-like. However, their assumed SKA1 sky fraction 
is 0.72, which is larger than our 0.48, which follows \citet{Bacon:2018dui}. 
Their LSST-like sky fraction is also chosen as 0.72, larger than our sky fraction 
for LSST-like of $\sim 0.33$, according to \citet{Mandelbaum:2018ouv}. 
The bias fitting functions used are the same as ours, and the same kind of nuisance parameters are introduced.
The main driver of the difference in results from ours is again the greater sky area and smaller $\ell_{\rm min}$ that they assumed.

In \citet{Schmittfull:2017ffw}, the case of LSST-like clustering and CMB-S4 lensing in cross 
correlation is investigated. The uncertainties found are $\sigma\left( f_{\rm  NL} \right) \simeq 0.4$ or $\sigma\left( f_{\rm  NL} \right) \simeq 1.0$ 
for the cases where the minimum multipole for both tracers is either 2 or 20. 
The galaxy redshift distribution is split into 6 bins, extending over a larger redshift 
range $0<z<7$ and assuming  50 sources/arcmin$^2$. 
The sky fractions they used  are 0.5 for both CMB-S4 and LSST-like, assuming 100\% overlap.
Their fiducial bias  model is $b(z) = 1+z$ as opposed to the one we use, 
$b(z) = 0.95/D(z)$. 
Once again, the greater sky area and smaller $\ell_{\rm min}$ that they assumed produce more optimistic constraints than ours.  The larger redshift range that they considered is not as important.
If we use the assumptions made by them, we recover their results.

\section{Conclusions}
\label{sec:conc}

In this paper we have shown how up to three tracers of the cosmic density field can be used 
to extract precise measurements of perturbations on scales beyond the equality scale.  
Specifically we forecast that a conservative combination of an SKA1-MID HI intensity mapping survey  with 
the galaxy clustering from two 
photometric galaxy surveys (Euclid- and LSST-like), and with CMB lensing from CMB-S4, could reach uncertainties 
for primordial non-Gaussianity parameters of $\sigma(f_{\rm NL}) \lesssim 0.9$ and $\sigma(n_{\rm NL}) \lesssim 0.2$.
We highlighted the importance of CMB lensing information through the cross-correlation with 
intensity/ number counts to further improve the uncertainties on $f_{\rm NL}$.

The uncertainties obtained for local type PNG in the single-tracer cases are 
$\sigma(f_{\rm NL}) \simeq 2.1$ for SKA1-MID IM with $\ell_{\rm min} = 5$, 
$\sigma(f_{\rm NL}) \simeq 2.3$ for Euclid-like with $\ell_{\rm min} = 10$, 
and $\sigma(f_{\rm NL}) \simeq 16.2$ for LSST-like with $\ell_{\rm min} = 20$.  
On the running index of $f_{\rm NL}$ in the extended local PNG model, we found $\sigma(n_{\rm NL}) \simeq 2.7$, $0.35$, $0.37$
respectively.

Combining two different large-scale structure surveys via the multi-tracer approach, we forecast
 $\sigma(f_{\rm NL}) \simeq 0.96\ (1.6)$ for SKA1-MID IM 
with Euclid-like (LSST-like) and $\sigma(n_{\rm NL}) \simeq 0.13\ (0.24)$.

When we combine CMB lensing information (with $\ell_{\rm min} = 30$) from a possible CMB-S4 ground-based experiment in the multi-tracer, with a single LSS survey, we found that the single-tracer
errors decrease to $\sigma(f_{\rm NL}) \simeq 1.6$, $1.8$, $10.5$ for SKA1-HI
IM, Euclid-like, and LSST-like, respectively. 

When all three tracers are included in a multi-tracer analysis, 
the tightest uncertainties were predicted 
\begin{eqnarray}
&&{\sigma(f_{\rm NL}) \simeq 0.90~ \mbox{and}~ 
\sigma(n_{\rm NL}) \simeq 0.12}   \notag \\
&&{\mbox{for}~ \mbox{SKA1}\times \mbox{Euclid-like} \times \mbox{CMB-S4}.} 
\end{eqnarray}
Using LSST-like instead of Euclid-like, these degrade to $\sigma(f_{\rm NL}) \simeq 1.4$ 
and $\sigma(n_{\rm NL}) \simeq 0.22$.

We considered also the possibility of using simulated {\em Planck}-like data, leading to 
uncertainties on the cosmological parameters compatible with the latest results 
in \citet{Akrami:2018vks,Aghanim:2018eyx,Aghanim:2018oex} as representative of 
current CMB measurements.
In this case, the improvement in uncertainties by adding {\em Planck} to the single-tracer cases is very small and mostly due to parameter degeneracy with the standard 
cosmological parameters, rather than an imprinting of $f_{\rm NL}$ on the 
cross-correlation between intensity/number counts with CMB lensing.
We also tested the possibility of completing the missing first multipoles 
$2 \leq \ell_{\rm min} < 30$ in the CMB spectra, but we found no further improvement.

Constraints on PNG parameters from the measurement of ultra-large scales depend 
strongly on the $\ell_{\rm min}$ and $f_{\rm sky}$ considered in the analysis.
Our constraints use more conservative estimates and the most up-to-date specifications for the surveys involved. 
In light of the differences in assumptions made in  previous papers, 
it is not unexpected that our constraints are weaker.

We assumed the minimum multipoles and sky areas for each experiment according to 
up-to-date specifications for each survey:\\
$\ell_{\rm min}=~\,5,\,\Omega= 20,000\,$deg$^2$ -- SKA1 
\citep{Bacon:2018dui};\\ $\ell_{\rm min}=10,\,\Omega= 15,000\,$deg$^2$ -- Euclid-like \citep{Amendola:2016saw};\\ 
$\ell_{\rm min}=20,\,\Omega= 13,800\,$deg$^2$ -- LSST-like \citep{Mandelbaum:2018ouv};\\ 
$\ell_{\rm min}=30,\,\Omega= 16,500\,$deg$^2$ -- CMB-S4 \citep{Abazajian:2016yjj}.

We also studied how uncertainties change as a function of the
minimum multipole, shown in Fig.~\ref{fig:fNL_lmin}. 
For the multi-tracer sky overlap area, we took the smallest of the  sky fractions involved. 
For smaller overlaps, the uncertainties will be mildly negatively affected.

Finally, many other different tracers have been highlighted as good candidates to obtain 
competitive constraints on $f_{\rm NL}$, such as 
clusters of galaxies \citep{Pillepich:2011zz,Mana:2013qba,Sartoris:2015aga}, 
cosmic infrared background \citep{Tucci:2016hng}, 
cosmic voids \citep{Chan:2018piq} and different IM lines, like H$\alpha$, CO and CII
\citep{Fonseca:2018hsu,MoradinezhadDizgah:2018lac}. These could also be included in the analysis 
in order to reach more robust and tighter constraints.

\section*{Acknowledgements}

The authors were supported by the South African Radio Astronomy Observatory, which is a facility of 
the National Research Foundation, an agency of the Department of Science \& Technology. 
MB was also supported by the Claude Leon Foundation and by ASI n.I/023/12/0``Attivit\'a relative alla
fase B2/C per la missione Euclid". RM was also supported by the UK Science \& Technology 
Facilities Council (Grant no. ST/N000668/1).


\bibliographystyle{mnras}
\bibliography{Biblio}


\end{document}